# Fast and efficient critical state modelling of field-cooled bulk high-temperature superconductors using a backward computation method

Kai Zhang[1], Mark Ainslie[2], Marco Calvi[1], Sebastian Hellmann[1], Ryota Kinjo[3], Thomas Schmidt[1]

[1]Insertion Device Group, Photon Science Division, Paul Scherrer Institute, Villigen, 5232, Switzerland

[2]Bulk Superconductivity Group, Department of Engineering, University of Cambridge, Cambridge, CB2 1PZ, United Kingdom

[3]Research and Development Division, RIKEN SPring-8 Center, Hyogo, 679-5148, Japan

E-mail: kai.zhang@psi.ch

**Abstract**

A backward computation method has been developed to accelerate modelling of the critical state magnetization current in a staggered-array bulk high-temperature superconducting (HTS) undulator. The key concept is as follows: i) a large magnetization current is first generated on the surface of the HTS bulks after rapid field-cooling (FC) magnetization; ii) the magnetization current then relaxes inwards step-by-step obeying the critical state model; iii) after tens of backward iterations the magnetization current reaches a steady state. The simulation results show excellent agreement with the *H*-formulation method for both the electromagnetic and electromagnetic-mechanical coupled analyses, but with significantly faster computation speed. Solving the FEA model with 1.8 million degrees of freedom (DOFs), the backward computation method takes less than 1.4 hours, an order of magnitude or higher faster than other state-of-the-art numerical methods. Finally, the models are used to investigate the influence of the mechanical stress on the distribution of the critical state magnetization current and the undulator field along the central axis.

______________________________________________________________________________________

## 1. Introduction

Research and development work on short-period and high-field staggered-array high-temperature superconducting (HTS) bulk undulators [1,2] is ongoing in a European project for the construction of compact free electron lasers (FELs) [3,4]. This new technology utilizes a 10 T level superconducting solenoid magnet to realize field-cooling (FC) magnetization of a series of staggered-array ReBCO bulks at a temperature around 10 K. With this concept, a sinusoidal undulator field of amplitude 2 T with a period length of 10 mm along the central beam axis can be obtained [2,5]. One key challenge for this technology is that the mechanical properties of ReBCO bulk superconductors are ceramic-like: friendly towards compressive stress and unfriendly towards tensile stress. Thus, some form of external mechanical reinforcement is usually used to compress the ReBCO bulk to trap high magnetic fields [6-9]. Trillaud *et al*. (2018) showed the critical current density $J_c$ of ReBCO bulk superconductor will degrade when the Lorentz force-induced mechanical stress is of the order of the fracture strength [10]. This indicates that the critical current density $J_c$ should be a function of both the magnetic flux density ***B*** and the mechanical strain $\varepsilon$ when a ReBCO bulk superconductor traps a high magnetic field and experiences the associated large Lorentz force. Regarding the short-period and high-field staggered-array HTS bulk undulator, estimation of the magnetization current that follows a $J_c(\boldsymbol{B},\varepsilon)$-determined critical state model [11,12], without time-dependent flux creep effects, is of great interest for the purpose of optimizing the first and the second integrals of the undulator field along the central axis.

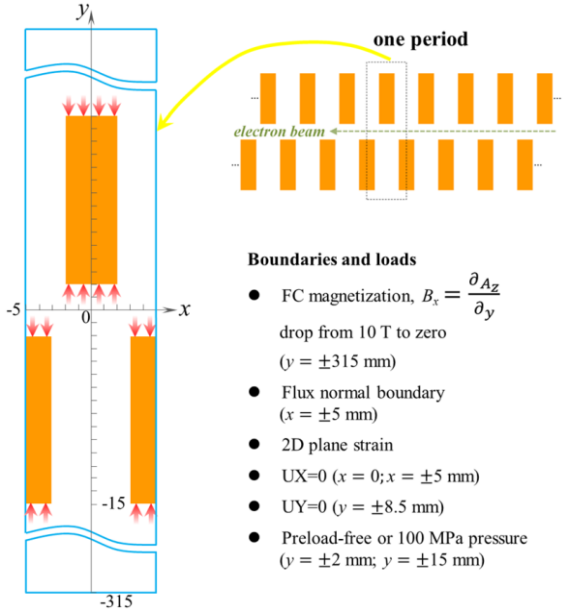
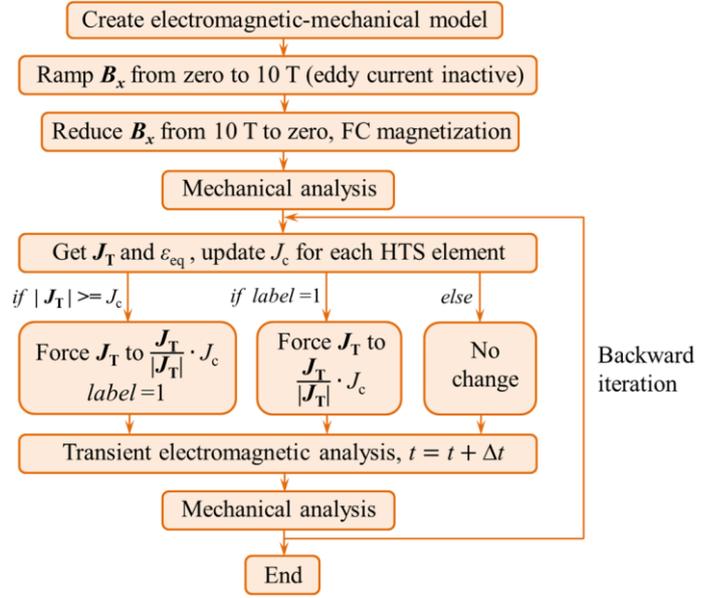

Figure 1. Periodical FEA model of a staggered array ReBCO bulk undulator with a period length of 10 mm and magnetic gap of 4 mm.

Figure 2. Backward computation of the critical state magnetization current after FC magnetization from 10 T. The trapped current density obeys to modified critical state model, $J_c(B,\varepsilon_{eq})$.

There are two main methods to compute the critical state model for type-II superconductors. Some methods calculate the critical current density directly by using complex numerical methods [13-15]. Others define an $\boldsymbol{E}$-$\boldsymbol{J}$ power law [16] or a flux-flow resistivity [17] in commercial FEA software like COMSOL [18,19], FlexPDE [17], GetDP [20], or Flux2D/3D [21]. Recently an iterative algorithm method was proposed to compute a $J_c(\boldsymbol{B},\theta)$-determined critical state model for ReBCO tape stacks [22]. It avoids using unnecessary iterative steps to obtain a resistivity matrix [23,24] but still requires hundreds of iterative steps to obtain adequate results. This paper introduces a new backward computation method to accelerate modelling the $J_c(\boldsymbol{B},\varepsilon)$-determined critical state magnetization current in the periodical HTS bulk undulator. It takes only tens of backward iterations to reach a steady-state solution, which can be an order of magnitude or higher faster than other state-of-the-art numerical methods.

## 2. FEA model and backward computation

Figure 1 shows the one-period 2D FEA model of the periodical HTS bulk undulator created in ANSYS 18.1 Academic. For the electromagnetic analysis, the magnetic flux density $\boldsymbol{B_x}$ is applied to the outer air subdomain boundaries @$y = \pm 315$ mm to provide the background magnetic field; a flux normal boundary (default in ANSYS) is applied to the boundaries on the sides @$x = \pm 315$ mm to model the periodicity/symmetry. For the mechanical analysis, a displacement constraint is applied the x-direction @$x = 0$ and $x = \pm 5$ mm and the y-direction @$y = \pm 8.5$ mm to avoid movement due to the action of the Lorentz force. The pre-stress, if any, is applied to the top and bottom sides of the HTS bulks as mechanical reinforcement to compensate the expected large tensile Lorentz force.

Figure 2 describes the algorithm of the backward computation. The background $B_x$ is first ramped from zero to 10 T (step 1) and then reduced from 10 T to zero over a short time (50 seconds; step 2). The eddy current solver is turned off during the first step and turned on during the second step to calculate the induced magnetization current. The resulting mechanical stress is analyzed by importing the Lorentz force and applying a pre-stress. Afterwards, a backward loop computation of the relaxation of the magnetization current is carried out as follows

a) Obtain the magnetization current $J_T$ and the equivalent mechanical strain $\varepsilon_{eq}$ for each HTS element, and update $J_c$;

b) For each HTS element, force the magnetization current $J_T$ to $J_c \cdot J_T/|J_T|$ if $|J_T| > J_c$ or the element has been penetrated (Each HTS element has a "*label*" with the default value of zero; once the HTS element is penetrated its "*label*" becomes 1);

c) Carry out the transient electromagnetic analysis with a small time increment ($\Delta t = 0.5$ s);

d) Carry out the 2D plane strain mechanical analysis.

During the backward iterations the resistivity of the superconductor is set to a fixed low value ($1 \times 10^{-15}$ $\Omega$m) and the *A-V* formulation is used for fast and efficient electromagnetic analysis.

$$\nabla \times A = B \tag{1}$$

$$\nabla \times B = \mu J \tag{2}$$

$$J = -\frac{1}{\rho}\left(\frac{\partial A}{\partial t} + \nabla V\right) \tag{3}$$

$$\nabla \times \left(\frac{1}{\mu}\nabla \times A\right) = -\frac{1}{\rho}\left(\frac{\partial A}{\partial t} + \nabla V\right) \tag{4}$$

The entire process follows these Maxwell's equations and the modified critical state model for which $J_c$ @10 K is expressed as

$$J_c(B, \varepsilon_{eq}) = k_{c,m}\left\{J_{c1}\exp\left(-\frac{B}{B_L}\right) + J_{c2}\frac{B}{B_{max}}\exp\left[\frac{1}{y}\left(1 - \left(\frac{B}{B_{max}}\right)^y\right)\right]\right\} \tag{5}$$

where $k_{c,m}$ is the mechanical degradation factor describing the $J_c$ degradation due to the mechanical stress. The assumed values of $J_{c1}$, $J_{c1}$, $B_L$, $B_{max}$ and $y$ are $1.0 \times 10^{10}$ A/m$^2$, $8.8 \times 10^9$ A/m$^2$, 0.8 T, 4.2 T and 0.8, respectively. These values refer to the $J_c$ data [25] of the ReBCO bulk @ 40 K and are scaled to 10 K from the first experimental result of our ten-period GdBCO bulk undulator tested at the University of Cambridge [2]. The mechanical degradation factor [10] is a function of the equivalent mechanical strain $\varepsilon_{eq}$

$$k_{c,m} = \left(1 - \gamma\left(\frac{\varepsilon_{eq}}{\varepsilon_c}\right)^2\right) \times \left[\alpha + \frac{1 - \alpha}{1 + \exp\left((|\varepsilon_{eq}/\varepsilon_c| - 1)/\beta\right)}\right] \tag{6}$$

where $\gamma = 0.1$, $\beta = 0.025$, $\alpha = 10\%$ and $\varepsilon_c = 6.0 \times 10^{-4}$ ($\sigma_c = 90$ MPa, $E = 150$ GPa).

## 3. Results and discussion

### 3.1. Computational results from the backward computation method

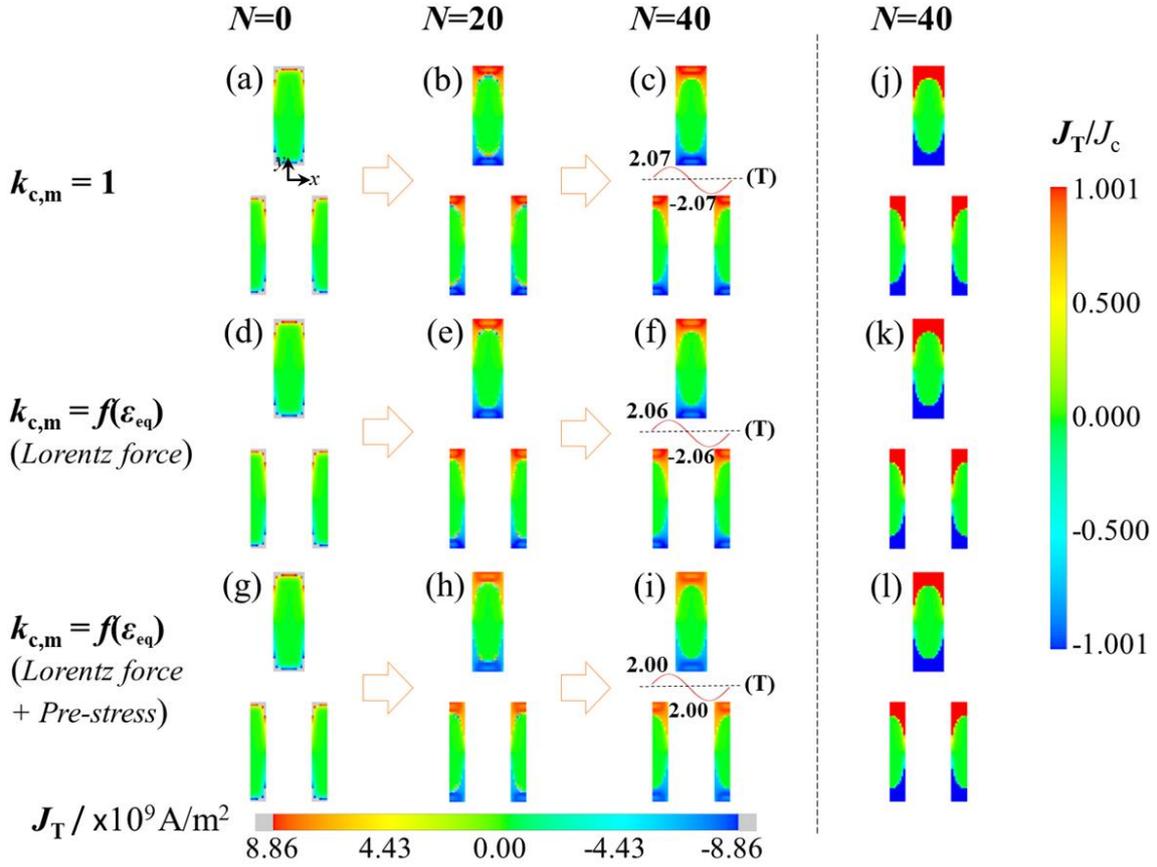

Figure 3. Magnetization current $J_T$ in the periodical HTS bulk undulator during the backward iterations (a) $N$=0, (b) $N$=20 and (c) $N$=40 without considering the mechanical degradation factor. Magnetization current $J_T$ during the backward iterations (d) $N$=0, (e) $N$=20 and (f) $N$=40 when considering the mechanical degradation factor due to the Lorentz force. Magnetization current $J_T$ during the backward iterations (g) $N$=0, (h) $N$=20 and (i) $N$=40 when considering the mechanical degradation factor due to both the Lorentz force and the pre-stress. On the right (j-l) is $J_T$ normalized to $J_c$ for the three different cases. A stable sinusoidal magnetic field $B_y$ along the $x$-axis is generated after ~35 backward iterations.

The simulation results for three different cases are compared and plotted in figure 3 (multimedia view: load step 1-5, ramp $B_x$ from zero to 10 T; load step 6, damp $B_x$ from 10 T to zero; load step 7, start backward iterations). In figure 3(a) the outermost layer of the HTS bulks traps large magnetization current $J_T$ (gray color) after the quick FC magnetization in 50 s. Figures 3(b) and (c) show the magnetization current $J_T$ in the HTS bulk undulator after 20 and 40 backward iterations, respectively. In fact, after ~35 iterations the magnetization current no longer changes and the induced sinusoidal undulator field $B_y$ along the $x$-axis becomes stable with an amplitude of 2.07 T. Figures 3(d)-(f) show the evolution of the magnetization current $J_T$ after considering the mechanical degradation factor $k_{c,m}$ resulting from the Lorentz force. The magnetization current $J_T$ also becomes stable after ~35 iterations. The induced sinusoidal undulator field $B_y$ along the $x$-axis now has an amplitude of 2.06 T. Figures 3(g)-(i) show the evolution of the magnetization current $J_T$ after considering the mechanical degradation factor $k_{c,m}$ resulting from the Lorentz force and the pre-stress. After ~35 backward iterations the induced sinusoidal undulator field $B_y$ becomes stable but with a much lower amplitude of 2.00 T. This phenomenon can be explained by the $J_c$ reduction in the outer layer of the HTS bulks due to the non-negligible von Mises stress. In other words, the third case (Lorentz force + pre-stress) has the most penetrated HTS elements but the lowest averaged $\overline{|J_T|}$ in the penetrated region. All three solutions retain a magnetic flux density $B_x$ of 10 T in the unpenetrated HTS region. Figures 3(j)-(l) show the $J_T$ normalized to $J_c$ for the three cases. The values are ±1 for the penetrated HTS elements.

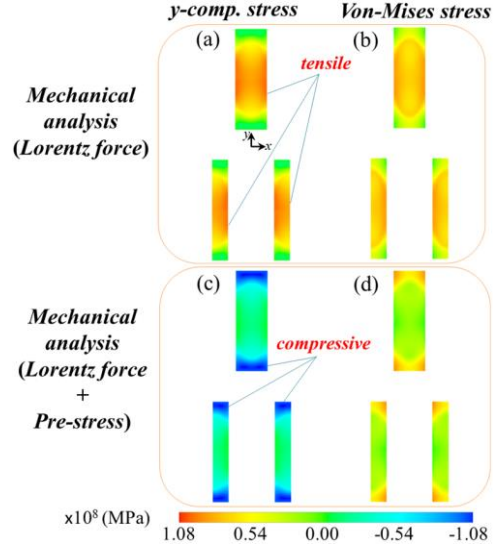

Figure 4. Mechanical stress in the periodical HTS bulk undulator after FC magnetization from 10 T. The tensile stress in the *y*-direction becomes compressive after applying the pre-stress.

Figures 4(a) and (b) show the mechanical stress ($\sigma_y$, the y-component stress, and $\sigma_v$, the von Mises stress) due to the Lorentz force only after the FC magnetization from 10 T to zero. The *y*-component stress is tensile and around 100 MPa in the bulk center, which is unacceptable for the brittle ceramic material. Figures 4(c) and (d) show the mechanical stress ($\sigma_y$ and $\sigma_v$) due to the Lorentz force and the 100 MPa pre-stress. It can be observed that the Lorentz force-induced tensile stress in the *y*-direction is compensated. In the meantime, the high von Mises stress region shifts from the bulk center to the bulk ends. This explains the $J_c$ reduction in the outer layer of the bulk HTS. In both of the two simulation cases, the peak von Mises stress in the HTS bulks is around 90 MPa, but the latter case exhibits a compressive stress in all three main directions, much less detrimental to the ceramic-like bulk HTS material.

The simulation results confirm two facts: a) applying a pre-stress on the bulk HTS can enhance its mechanical performance for the purpose of trapping high magnetic field; b) the applied pre-stress can affect the distribution of the magnetization current in the bulk HTS, thus reducing the undulator field along the central axis.

## 3.2. Validation by the electromagnetic-mechanical coupled *H*-formulation

In this section, the electromagnetic properties of the bulks are simulated using the *H*-formulation, implemented in COMSOL Multiphysics (version 5.4) using the 'Magnetic Field Formulation' interface in COMSOL's AC/DC module. Both COMSOL and the *H*-formulation are currently used by dozens of research groups worldwide to model bulk superconductors [26,27] and other superconductivity-related problems [28,29].

For the 2D *H*-formulation, the independent variables are the components of the magnetic field strength, $H = [H_x\ H_y\ 0]$, and the governing equations are derived from the Maxwell's equations – namely, Ampere's (7) and Faraday's (8) laws:

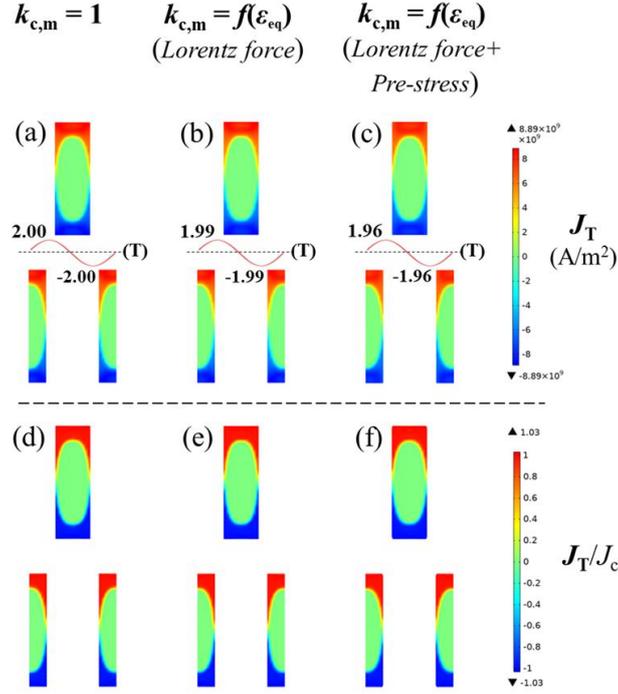

Figure 5. Magnetization simulation results using the mechanical-coupled *H*-formulation implemented in *COMSOL® 5.4*. ($n$ = 100). (a) Magnetization current $J_T$ without considering the mechanical degradation factor; (b) Magnetization current $J_T$ when considering the mechanical degradation factor due to the Lorentz force; (c) Magnetization current $J_T$ when considering the mechanical degradation factor due to the Lorentz force and the pre-stress. At the bottom (d-f) is $J_T$ normalized to $J_c$ for the three different cases.

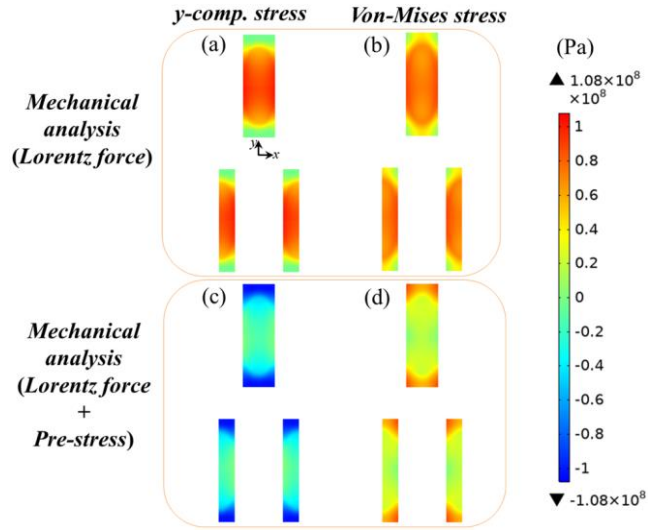

Figure 6. Mechanical stress obtained using the mechanical-coupled *H*-formulation in *COMSOL® 5.4*.

$$\nabla \times \boldsymbol{H} = \boldsymbol{J} \tag{7}$$

$$\nabla \times \boldsymbol{E} = -\frac{\partial \boldsymbol{B}}{\partial t} \tag{8}$$

The permeability $\mu = \mu_0$, and equations (7) and (8) are combined with the *E-J* power law (9), used to simulate the nonlinear resistivity, $\rho(J)$, of the superconductor [30-32]:

$$E = \frac{E_0}{J_c}\left|\frac{J}{J_c}\right|^{n-1} \quad (9)$$

$J = [0\ 0\ J_z]$ and $E = [0\ 0\ E_z]$ are the current density and electric field, respectively, which are assumed to be parallel to each other such that $E = \rho J$. $E_0 = 1\ \mu V/cm$ is the characteristic electric field and $n$ defines the steepness of the transition between the superconducting state and the normal state; we assume here that $n = 100$ to reasonably approximate the critical state model [27,33].

FC magnetization is simulated by setting an appropriate magnetic field boundary condition to the top and bottom outer boundary conditions such that, for $0 \leq t \leq 100$ s, $H_x(t) = 10 - t/t_{ramp}$, where $t_{ramp} = 10$ s. Thus, we have the initial condition, $H_x(t = 0\ s) = 10$ T and the magnetic field is ramped linearly down to $H_x(t = 100\ s) = 0$ T. On the left- and right-side boundaries, the 'Perfect Magnetic Boundary' node ($n \times H = 0$) is used to model periodicity/symmetry. Since no net transport current flows, a constraint is applied to each of the bulks such that, at all times:

$$I(t) = \iint_S J_z\ dS = 0 \quad (10)$$

Isothermal conditions are assumed while ramping down the field, so no thermal model is included.

The electromagnetic model is coupled with COMSOL's 'Solid Mechanics' interface as described in [34]. The Lorentz force, $F_L = J \times B$, is implemented as a force per unit volume using the 'Body Load' node, where $F_x = -J_z \cdot B_y$ and $F_y = J_z \cdot B_x$. The displacement constraints are added using the 'Prescribed Displacement' node. The pre-stress is applied using the 'Boundary Load' node such that $F_{pre} = -pn$, where $p$ is the applied pressure.

Figure 5 shows the magnetization simulation results at "$t= 100$ s" for the three cases above. In figures 5(a)-(c) we can observe the peak magnetization current $J_T$ is $8.89 \times 10^9$ A/m$^2$, extremely close to the peak value of $8.86 \times 10^9$ A/m$^2$ shown in figure 3. In figures 5(d)-(f) we can observe the normalized $J_T$ to $J_c$ is $\sim\pm1$. The peak value is $\pm1.03$ which suggests a small flux creep effect exists, due to the finite (but high) $n$ value used. The induced sinusoidal undulator field $B_y$ along the x-axis has an amplitude of 2.00 T without considering the mechanical effects. The amplitude drops to 1.99 T and then 1.96 T when considering the Lorentz force and both the Lorentz force and the pre-stress, respectively. The slightly lower undulator field obtained by the COMSOL $H$-formulation can be explained by the unavoidable slight flux creep effect which can result in a lower averaged $\overline{|J_T|}$. Figure 6 shows the mechanical stress in the periodical HTS bulk undulator at "$t= 100$ s". Overall, the simulation results are highly consistent with those shown in figure 4.

### 3.3. Discussion

The backward computation method has been proven successful to model the critical state magnetization current in the HTS bulk undulator after FC magnetization. Compared to the $H$-formulation it has several advantages for the electromagnetic modelling of superconductors:

(a) The 2D $H$-formulation has degrees of freedom (DOFs) for $H_x$ and $H_y$ for the entire FEA model; however, the backward computation method uses the $A$-$V$ formulation which requires much lower number of DOFs ($A_Z$ and $V$ in the HTS subdomains, $A_Z$ in the air subdomain).

Table 1. Comparison of the number of DOFs and computation times between the two different methods

|  | **H**-formulation | Backward computation |
|---|---|---|
| No. of HTS elements | 1664 | 1664 |
| No. of total elements | 100800 | 100800 |
| No. of HTS DOFs | 3610 | 10742 |
| No. of total DOFs | 207870 | 114338 |
| Computation time (electromagnetic, EM) | 14 min | 3.5 min |
| Computation time (EM-mechanical) | 48 min | 3.5 min |

Table 2. Comparison of the number of DOFs and computation time for the backward computation

|  | Element size (0.125 mm x 0.125 mm) | Element size (0.0625 mm x 0.0625 mm) |
|---|---|---|
| No. of HTS elements | 6656 | 26624 |
| No. of total elements | 403200 | 1613368 |
| No. of HTS DOFs | 41446 | 162758 |
| No. of total DOFs | 452034 | 1799210 |
| Computation time (electromagnetic, EM) | 15 min | 79 min |
| Computation time (EM-mechanical) | 15 min | 82 min |

(b) The backward computation method does not solve the eddy current in the air region, thus reducing the computation time.

(c) The backward computation method solves the eddy current in the HTS subdomains by defining a fixed low resistivity. Solving an equation representing the nonlinear resistivity, such as the **E-J** power law, is not required.

In order to demonstrate the high efficiency of the backward computation method, we conducted two identical simulations using the COMSOL **H**-formulation and using the ANSYS backward computation. The same number of meshing elements is achieved by using mapped meshing (with element size: 0.25 mm x 0.25 mm) for the whole FEA model. In the **H**-formulation, the whole FEA model consisted of linear (first-order) quadrilateral elements. In the backward computation, the HTS and its surrounding air region consisted of second-order quadrilateral elements to obtain accurate solution results in this local region and the rest of the air region consisted of first-order quadrilateral elements. The number of mesh elements and DOFs and the computation times are listed in Table 1. For the electromagnetic-only analysis, the backward computation is ~4 times faster than the **H**-formulation; for running the electromagnetic-mechanical coupled analysis, the backward computation is ~14 times faster than the **H**-formulation. The rapid coupling analysis suggests the backward computation can solve any modified critical state models without reducing computation speed. It should be pointed out that the COMSOL **H**-formulation was run on a PC with an

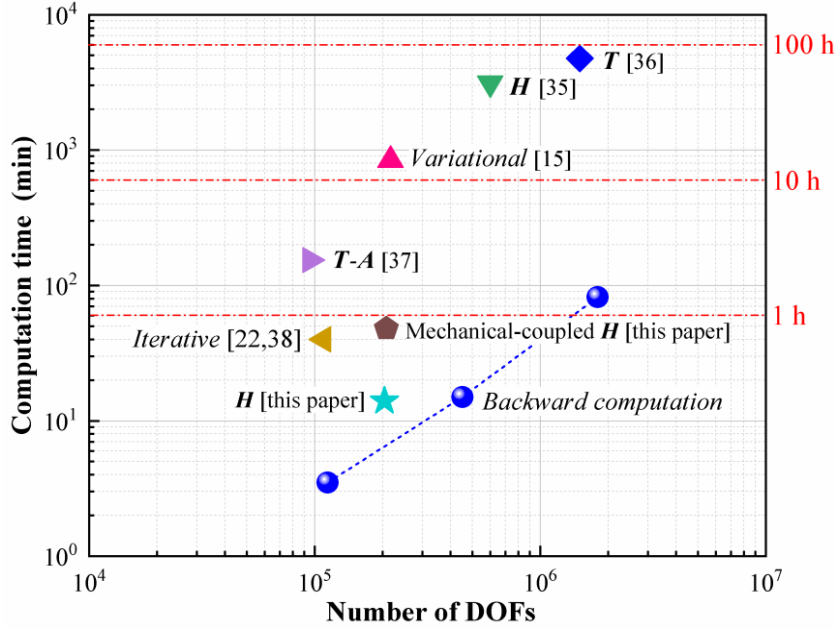

Figure 7. Comparison of computation times reported in the literature for other state-of-the-art techniques for the electromagnetic analysis of HTS materials.

Intel® Core™ i9-7900X CPU @ 3.3 GHz and 64 GB RAM, and the ANSYS backward computation was run on a PC with an Intel® Xeon® CPU E3-1245 v6 @ 3.7 GHz and 64 GB RAM. There is room for the computation speed of the backward computation to be improved with a higher-performance CPU.

To better understand the efficiency of the backward computation, we conducted two additional simulations by reducing the element size (increasing the total number of DOFs) in the HTS undulator model. As shown in Table 2, the total number of DOFs increases to 452,034 and then 1,799,210 when the element size is "0.125 mm x 0.125 mm" and "0.0625 mm x 0.0625 mm", respectively. It took approximately 15 minutes and 82 minutes, respectively, to run the electromagnetic-mechanical coupled simulation. We have compared these results with other state-of-the-art techniques for the electromagnetic analysis of HTS materials, like the ***H***-formulation [35], the ***T***-formulation [36], the variational method [15], the ***T-A*** formulation [37], and the recently proposed iterative algorithm method [22,38]. As shown in figure 7, the backward computation shows a surprising order of magnitude computation speed than all the other methods. It should be noted, however, that the listed ***H***-formulation, ***T***-formulation and ***T-A*** formulation were implemented for other applications (e.g., AC loss or screening-current-induced fields) and that benchmarking this particular problem would provide a true comparison. Nevertheless, solving such a large-scale HTS electromagnetic problem with 1.8 million DOFs within 1.4 hours is remarkably fast and was achieved using a normal PC.

## 4. Conclusion

We have demonstrated that the backward computation method can model the critical state magnetization current in a staggered array HTS bulk undulator quickly and efficiently by running benchmark simulations using a mechanical-coupled ***H***-formulation in COMSOL. The algorithm of the backward iterations is realized by utilizing the function of multi-frame restart analysis and the ***A-V*** formulation in ANSYS 18.1 Academic. The highly efficient computation, even with millions of DOFs, is because a nonlinear resistivity equation is not required and no eddy current is solved in the non-superconductor regions. These advantages, along with the backward concept itself, make this new method superior to many other numerical methods used to model the critical state magnetization current. Finally, we show that applying a pre-stress to the HTS bulks could enhance their mechanical performance when trapping high magnetic fields, but could result in a reduction in $J_c$ in the outer layer of the HTS bulks, thus reducing the induced undulator field. This important information will help guide future optimization of the integral undulator field along the meters-long central axis.

## Acknowledgement

This work is supported by European Union's Horizon2020 research and innovation program under grant agreement No 777431. Dr. Mark Ainslie would like to acknowledge financial support from an Engineering and Physical Sciences Research Council (EPSRC) Early Career Fellowship EP/P020313/1. All data are provided in full in the results section of this paper.